\documentclass[trackchanges,twocolumn]{aastex701}
\usepackage{amsmath}
\usepackage{pifont}
\usepackage{mathrsfs}

\begin{document}

\title{\vspace*{-.2in} Diffusion-Free Dynamics in Rotating Spherical Shell Convection \\ Driven By Internal Heating and Cooling}

\author[orcid=0000-0012-3245-1234]{Neil T. Lewis}
\affiliation{Department of Physics and Astronomy, University of Exeter, UK}
\email[show]{n.t.lewis@exeter.ac.uk}  

\author[orcid=0000-0002-8593-3461]{Tom Joshi--Hartley} 
\affiliation{Department of Physics and Astronomy, University of Exeter, UK}
\email{tj294@exeter.ac.uk}

\author[orcid=0000-0003-0205-7716]{Steven M. Tobias}
\affiliation{School of Physics and Astronomy, University of Edinburgh, UK}
\email{s.tobias@ed.ac.uk}

\author[orcid=0000-0002-0865-3861]{Laura K. Currie}
\affiliation{Department of Mathematical Sciences, Durham University, UK}
\email{laura.currie@durham.ac.uk}

\author[orcid=0000-0002-8634-1003]{Matthew K. Browning}
\affiliation{Department of Physics and Astronomy, University of Exeter, UK}
\email{m.k.m.browning@exeter.ac.uk}

\begin{abstract} The bulk properties of convection in stellar and giant planet interiors are often assumed to be independent of the molecular diffusivities, which are very small. By contrast, simulations of this process in rotating, spherical shells, which are typically driven by conductive boundary heat fluxes, generally yield results that depend on the diffusivity. This makes it challenging to extrapolate these simulation results to real objects. However, laboratory models and Cartesian-box simulations suggest that diffusion-free dynamics are more readily obtained if convection is driven using prescribed internal heating and cooling instead of boundary fluxes. Here, we apply this methodology to simulations of Boussinesq, hydrodynamic rotating spherical shell convection. We find that this set-up unambiguously yields diffusion-free behaviour for some bulk properties of the convection, such as the radial temperature contrast and the convective heat transport. Moreover, the transition from prograde to retrograde equatorial zonal flow is diffusion-free and only depends on the convective Rossby number. The diffusivity dependence of other bulk properties is regime-dependent. In simulations that are rotationally constrained, the convective velocities, and the strength and structure of the zonal flow, are diffusion-dependent, although the zonal flow appears to approach a diffusion-free state for sufficiently high supercriticality. In simulations that are uninfluenced by rotation, or are only influenced by rotation at large scales, diffusion-free convective velocities and zonal flows are obtained. The result that many aspects of our idealised simulations are diffusion-free has promising implications for the development of realistic stellar and giant planet convection models that can access diffusion-free regimes.
\end{abstract}



\section{Introduction}\label{sec:intro}

Convection in stellar and giant planet interiors is extremely turbulent, which poses a practical barrier to numerical simulation. To retain computational feasibility, simulations of this process are typically conducted in a parameter regime far from reality \citep{2020ASSP...57...75B,2023SSRv..219...58K}, with the results subsequently extrapolated to real objects through scaling analysis and by appealing to theory \citep{2020PhRvR...2d3115A}. In order to achieve this, it is desirable that the simulations exhibit the same scaling behaviour as would a real planet or star. It is typically assumed that the bulk properties of the convection should not depend on the molecular diffusivities, which are small in real astrophysical environments \citep{2012PhRvL.109y4503J,2014ApJ...791...13B}. This assumption underpins the derivation of mixing length theory (MLT; \citealp{1958ZA.....46..108B}), which has been applied with noteworthy success as a parametrisation of convective heat transport in stellar evolution models  (e.g., \citealp{2015A&A...577A..42B}). 

However, diffusivity independent (i.e., `diffusion-free') dynamics are rarely obtained in numerical simulations. For example, the vast majority of shell convection simulations yield scaling behaviour for the radial heat transport that depends upon the diffusivity \citep{2002JFM...470..115C,2012Icar..219..428G,2015JFM...778..721G,2016JFM...808..690G,2016ApJ...818...32F,2016AdSpR..58.1475O}, although a notable exception is presented by \citet{2016JFM...808..690G}, where a subset of their simulations do obtain a diffusion-free scaling for cases with very rapid rotation (paired with moderate supercriticality). The influence of diffusivity in simulations of shell convection also affects their bulk kinematic properties. For example, simulations of stellar convection with high diffusivity have been shown to generate prograde, `Solar-like' equatorial zonal flows (or `differential rotation'), but they have a tendency to switch to a regime of retrograde, `anti-Solar' flow as the diffusivities are reduced \citep{2013Icar..225..156G,2014MNRAS.438L..76G,2014A&A...570A..43K,2015ApJ...804...67F,2020ApJ...898..120H}. Moreover, \citet{2016JFM...808..690G} obtain a diffusivity-dependent scaling  for the Reynolds number even for simulations that are highly supercritical.

Driving convection using a conductive heat flux through the inner and outer radial boundaries is one cause of diffusivity dependence in numerical simulations. When convection is boundary-driven, thin thermal boundary layers form, and the heat transport through the domain is throttled by the conductive flux through the boundary layer (\citealp{2016JFM...808..690G}; see also the review by \citealp{2009RvMP...81..503A}). Regarding the differential rotation, the transition from prograde to retrograde motion is understood to occur when the convective Rossby number, which for a fixed rotation rate generally increases  as the diffusivities are reduced (in simulations with diffusion-dependent dynamics), exceeds unity \citep{1977GApFD...8...93G,2014MNRAS.438L..76G}. \citet{2022ApJ...938...65C} provide a geometric interpretation for this transition, while other authors (e.g.,  \citealp{2020ApJ...898..120H}) argue it is directly related to the thinning of the thermal boundary layer. It is interesting to note that simulations of deep stellar convection driven entirely by internal heating and cooling also exhibit diffusivity-dependent differential rotation \citep{2002ApJ...570..865B,2008ApJ...673..557M,2008ApJ...676.1262B}; however, in these models conductive heat transport is likely still important near the upper boundary, owing to the thinness of the cooling layer. 

It can be argued `boundary-driven' simulations are not particularly relevant to real astrophysical environments, as the boundary layers that dominate the dynamics over large swathes of parameter space likely bear little resemblance to those of the real objects \citep{2014ApJ...791...13B,2020ApJ...898..120H}. This has led some authors to pursue an alternative approach, where convection is driven by radiative heating and cooling (with a specified spatial structure) applied directly to the fluid. In the context of rapidly rotating $f$-plane convection in a Cartesian geometry, \citet{2014ApJ...791...13B} demonstrate that this approach yields diffusion-free bulk scaling behaviour for both thermal (e.g., the convective heat transport) and kinematic (e.g., the fluctuating kinetic energy) properties. \citet{2020MNRAS.493.5233C} generalise this result to the case where the directions of the acceleration due to gravity and the rotational axis are misaligned (as is the case for the non-polar latitudes of a sphere). However, \citet{2025MNRAS.541.2291J} illustrate that it is possible to simultaneously obtain diffusion-free heat transport, but diffusion-dependent convective velocities (using simulations of internally heated and cooled, rotating convection in an idealised Cartesian geometry with $2.5$ dimensions). Additionally, they show that the realisation of diffusion-free dynamics in their model is sensitive to the kinematic boundary condition (free-slip configurations more readily yield diffusion-free behaviour). \citet{2022PhRvL.129b4501K} show that diffusion-free heat transport can be obtained in Cartesian simulations of non-rotating convection driven by internal heating and cooling (i.e., a realisation of the `ultimate regime'). Finally, laboratory experiments driven by internal heating and cooling have obtained diffusion-free heat transport for both the rotating case and the non-rotating case \citep{2018PNAS..115.8937L,2021PNAS..11805015B,2024JFM...998A...9H}.

Building on these studies, our objective is to investigate the dynamics and scaling behaviour of convection driven by internal heating and cooling in a spherical geometry. We aim to determine whether internal heating and cooling gives rise to `diffusion-free' dynamics, focusing on the convective heat transport, convective velocities, and the zonal flows (differential rotation).   We also investigate the extent to which our simulation results can be described by standard scaling theories (whether diffusion-free or otherwise). In this work, we  study the dynamics of a Boussinesq, hydrodynamic fluid, deferring inclusion of the effects of compressibility and magnetism to future work.

\section{Methods}\label{sec:methods}

We study the dynamics of Boussinesq, rotating spherical shell convection driven by internal heating and cooling. Our simulations are conducted using the pseudo-spectral code \emph{Dedalus} \citep{2020PhRvR...2b3068B}. All variables are expanded in terms of spherical harmonics in the horizontal directions and Chebyshev polynomials in the radial direction using a standard 3/2 dealiasing. 

\subsection{Model}

The governing equations are \begin{align}
    \frac{\text{D}\boldsymbol{u}}{\text{D}t} + 2\mathbf{\Omega}\times\boldsymbol{u} &= -\mathbf{\nabla}p + \tilde{g}(r)T\mathbf{e}_{r} + \nu\nabla^{2}\boldsymbol{u} \\ 
    \mathbf{\nabla}\cdot\boldsymbol{u}&=0\\ 
    \frac{\text{D}T}{\text{D}t} &= q(r) + \kappa\nabla^{2}T, 
\end{align}
where $\boldsymbol{u}=(u_{r}, u_{\theta}, u_{\phi})$ is the fluid velocity in spherical polar coordinates, $p$ is a pressure, and $T$ is a scaled temperature variable, related to the real temperature by $T = g\alpha \delta T_{\text{real}}$ (where $g$ is the magnitude of gravitational acceleration and $\alpha$ is the coefficient of thermal expansion, and $\delta T_{\text{real}}$ is the Boussinesq temperature perturbation). The symbol $\nu$ denotes the kinematic viscosity and $\kappa$ denotes the thermal diffusivity. The dimensionless gravity profile is given by $\tilde{g}(r)=\left(r_{\text{o}}/r\right)^{2}$ where $r_{\text{o}}$ is the outer radius \emph{of the convection zone} (see below for definition). The unit vector in the radial direction is given by $\mathbf{e}_{r}$, and the rotation vector $\mathbf{\Omega}=\Omega_{0}(\cos\theta, -\sin\theta,\ 0)$, where $\theta$ is co-latitude and $\Omega_{0}$ is the rotation rate. 

Convection is driven by a prescribed heating and cooling function $q(r)$, configured so that heat is deposited and removed from regions of depth $\delta$ at the base and top of the simulated domain, respectively. We use the notation $r_{\text{i}}$  and $r_{\text{o}}$ to denote the radii of the inner and outer boundaries \emph{of the convection zone} (CZ), within which there is no imposed heating or cooling. Mathematically, $q$ is defined (following \citealp{2014ApJ...791...13B,2020MNRAS.493.5233C}): \begin{equation} 
q(r)=\frac{F}{\delta}\frac{r_{\text{i}}^{2}}{r^{2}}\begin{cases}
1+\cos\left[\frac{2\pi\left(r-r_{\text{i}}+\frac{\delta}{2}\right)}{\delta}\right]\kern-3pt; & r_{\text{i}}-\delta\leq r\leq r_{\text{i}}, \\ 
0; & r_{\text{i}} < r < r_{\text{o}}, \\ 
-1-\cos\left[\frac{2\pi\left(r-r_{\text{o}}-\frac{\delta}{2}\right)}{\delta}\right]\kern-3pt; & r_{\text{o}}\leq r\leq r_{\text{o}}+\delta,
\end{cases}
\end{equation}
where $F$ is the flux injected at the base of the CZ (corresponding to a shell-integrated flux $\mathscr{F}_{\text{tot}}=4\pi r_{\text{i}}^{2}F$; $\mathscr{F}_{\text{tot}}$ is independent of $r$ within the CZ). The geometric prefactor in the definition of $q$ ensures that the cooling layer removes the same amount of heat as is injected by the heating layer. 

At the inner and outer radii of the simulated domain (located at $r_{\text{i}}-\delta$ and $r_{\text{o}}+\delta$, respectively) we apply a free-slip, impenetrable, and insulating boundary condition (consistent with $\iiint_{V} q(r)\,\text{d}V=0$). Applying the heating and cooling function $q(r)$ in combination with these boundary conditions has the effect of replacing the conductive boundary layers that exist in boundary driven simulations with heating and cooling regions of fixed size.

\subsection{Non-dimensionalisation}\label{sec:nd}

To non-dimensionalise the governing equations, we use the depth of the convection zone $d\equiv r_{\text{o}}-r_{\text{i}}$ as a reference length scale, a flux-based temperature scale, given by $\Lambda\equiv Fd/\kappa$, and a timescale given by the free-fall time $\tau=\sqrt{d/\Lambda}$. 

Under this non-dimensionalisation, the dynamics of the system are determined by: a flux-based Rayleigh number, $Ra_{\text{F}}$; the Taylor number, $Ta$; the Prandtl number, $Pr$; the CZ radius ratio, $\eta$; and the non-dimensional heating/cooling depth $\tilde{\delta}$. These numbers are defined by: \begin{equation}
\begin{split}
&Ra_{\text{F}}\equiv\frac{d^{4}F}{\nu\kappa^{2}},\quad Ta\equiv\frac{4\Omega_{0}^{2}d^{4}}{\nu^{2}},\quad Pr\equiv\frac{\nu}{\kappa},\\ &\eta\equiv\frac{r_{\text{i}}}{r_{\text{o}}},\ \text{and}\quad \tilde{\delta}\equiv\frac{\delta}{d}. 
\end{split}
\end{equation}
In all simulations, we set $Pr=1$, $\eta=0.8$, and $\tilde{\delta}=0.2$. From the definitions of $Ra_{\text{F}}$, $Ta$, and $Pr$, it is possible to define a flux-based convective Rossby number (e.g., \citealp{2002JFM...470..115C,2006GeoJI.166...97C,2020ApJ...898..120H,2024A&A...683A.221K}), given by: \begin{equation} 
    Ro_{\text{cv,F}}\equiv\left(\frac{Ra_{\text{F}}}{Ta^{\frac{3}{2}}Pr^{2}}\right)^{\frac{1}{3}}=\left(d F\right)^{\frac{1}{3}}\frac{1}{2\Omega_{0} d}. \label{eq:Rocv}
\end{equation}
which, notably, is independent of the diffusivities $\nu$ and $\kappa$. Therefore, for a fixed flux, variation of $Ro_{\text{cv,F}}$ corresponds to varying $\Omega_{0}$, while varying $Ra_\text{F}$ and $Ta$ so as to keep $Ro_{\text{cv,F}}$ fixed is equivalent to varying $\nu$ and $\kappa$ (recall $\kappa=\nu$ when $Pr=1$). We note that this definition of $Ro_{\text{cv,F}}$ is closely related to estimates of the Rossby number commonly employed in stellar astronomy (e.g., \citealp{1984ApJ...279..763N}), $Ro\,{\sim}\,1/(\Omega_{0}\tau_{\text{c}})$, assuming that the convective overturning time $\tau_{\text{c}}$ is given by MLT.

\subsection{Summary of experiments and numerical details}

\begin{figure*}[ht!]
\centering\includegraphics[width=0.925\textwidth]{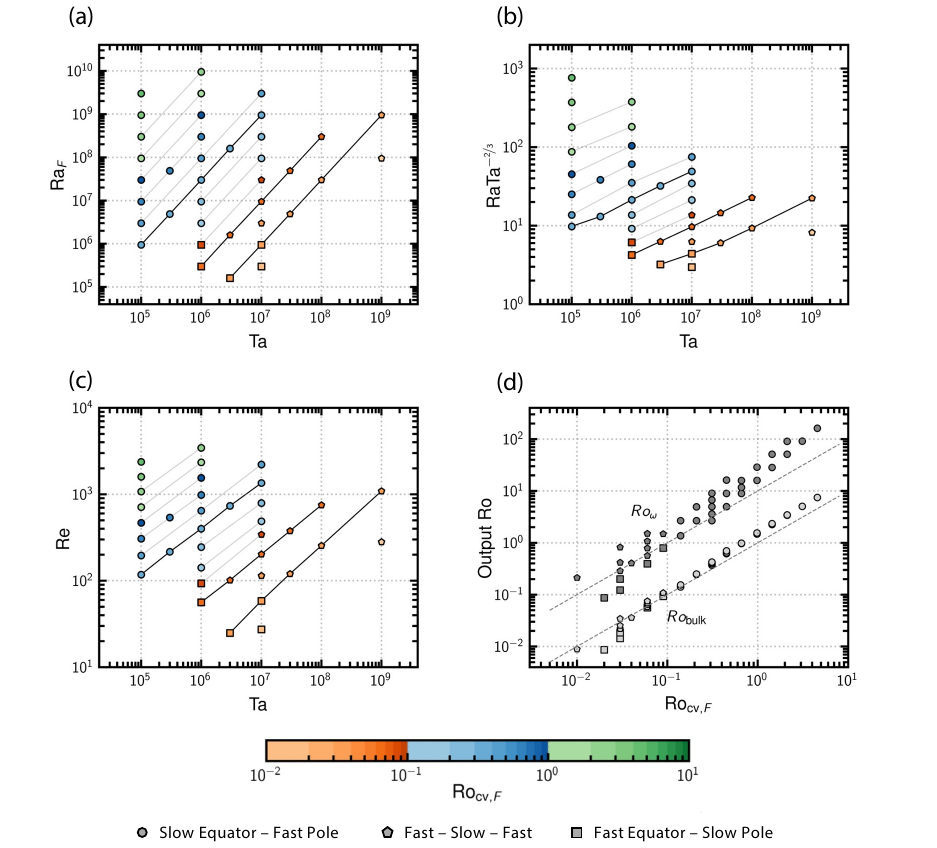}
\caption{Summary of simulations presented in this study. Panel a) shows the input parameters $Ra_{\text{F}}$ and $Ta$ used for each experiment. Panel b) shows the combination $Ra Ta^{-\frac{2}{3}}$ as a proxy for the supercriticality (computed using the `output' Rayleigh number given by Equation \ref{eq:Ra}). Panel c) shows $Re$ (defined by Equation \ref{eq:Re}), computed using the r.m.s. fluctuating velocity, as a measure of the degree of turbulence obtained in each simulation. Finally, panel d) shows the output Rossby numbers $Ro_{\text{bulk}}$ and $Ro_{\omega}$ (defined in Equation \ref{eq:Ro_alt}) as a measure of the degree of rotational constraint at the largest spatial scale, and smaller scales, respectively, plotted against $Ro_{\text{cv,F}}$. The marker colour denotes the dynamical regime occupied by a simulation, and the marker shape denotes the configuration of the differential rotation.  In panels a)--c), simulations that share the same $Ro_{\text{cv,F}}$ are connected by lines. Three lines are emphasised in bold, corresponding to $Ro_{\text{cv,F}}=3.11\times10^{-2}, 6.70\times10^{-2},\ \text{and}\ 3.11\times10^{-1}$ (from bottom to top in each panel). In panel d), the grey dashed lines are proportional to $Ro_{\text{cv,F}}$ and included as an eye guide. 
}\label{fig:param_summary}
\end{figure*}

Results from 38 simulations are presented. The parameter survey has been constructed to produce sets of simulations with constant $Ta$ (within which $Ra_{\text{F}}$, and thus $Ro_{\text{cv,F}}$, varies), as well as some sets with constant $Ro_{\text{cv,F}}$ (within which  $Ra_{\text{F}}$ and $Ta$ vary). The parameter survey spans a range of $Ro_{\text{cv,F}}$ between $0.01$ and $3.1$,  $Ta$ between $10^{5}$ and $10^{9}$, and $Ra_\text{F}$ between $3\times10^{5}$ and $9.5\times10^{9}$. A full list of our experiments is given in Appendix \ref{ap:A} (along with details of the resolution used), and a visual summary is shown in Figure \ref{fig:param_summary}. 

Figure \ref{fig:param_summary}a shows the input parameters $Ra_{\text{F}}$ and $Ta$, Figure \ref{fig:param_summary}b shows $Ra Ta^{-\frac{2}{3}}$ as a proxy for the supercriticality $Ra/Ra_{\text{c}}$ ($Ra_{\text{c}}\,{\sim}\,Ta^{\frac{2}{3}}$ for large $Ta$; \citealp{1961hhs..book.....C}), and Figure \ref{fig:param_summary}c shows the output Reynolds number to illustrate the degree of turbulence in each experiment. Above, $Ra$ is the usual Rayleigh number, which is an output parameter for a fixed-flux set-up. It is defined as \begin{equation} 
Ra \equiv \frac{d^{3}\Delta T}{\nu\kappa},  \label{eq:Ra}
\end{equation}
where $\Delta T=T(r\,{=}\,r_{\text{o}}) - T(r\,{=}\,r_{\text{i}})$ is the output radial temperature contrast, evaluated between the boundaries of the CZ. The Reynolds number is defined \begin{equation}
Re \equiv \frac{UL}{\nu} \label{eq:Re}
\end{equation} 
where we take $U$ to be the root-mean-square (r.m.s.) of the output fluctuating (azimuthal average removed) velocity vector within the CZ, and $L=d$. Also shown in Figure \ref{fig:param_summary}d are two different output Rossby numbers, defined \begin{equation} 
Ro_{\text{bulk}} = \frac{U}{\Omega_{0} d}\quad\text{and}\quad Ro_\omega = \frac{\omega}{\Omega_{0}}, \label{eq:Ro_alt} \end{equation}
where for $Ro_{\omega}$, $\omega$ is the r.m.s fluctuating vorticity within the CZ. These two output $Ro$ are intended to indicate the degree of rotational constraint at the largest spatial scales, and at small scales, respectively (e.g., \citealp{2021PNAS..11822518V}). We note that $Ro_{\omega}$ returns very similar values to the `local Rossby number' $Ro_{l}=Ro_{\text{bulk}}\frac{\ell}{\pi}$, where $\ell$ is the mean spherical harmonic degree obtained from the (mid-plane) kinetic energy spectrum (\citealp{2014MNRAS.438L..76G}; not shown).

Scaling results obtained from our experiments will show that it is useful to categorise them in terms of three regimes: (1) a `rotationally-constrained' regime defined by $Ro_{\text{cv,F}}<0.1$, for which \emph{all} spatial scales are rotationally constrained (i.e., $Ro_{\omega}\lesssim1$ and $Ro_{\text{bulk}}\ll1$); (2) a `rotationally-influenced' regime defined by $0.1\leq Ro_{\text{cv,F}}<1$, where the largest scales are rotationally constrained, but (at least some) smaller scales are not ($Ro_{\omega}\gtrsim1$ and $Ro_{\text{bulk}}<1$); and (3) a `rotationally-uninfluenced' regime defined by $Ro_{\text{cv,F}}\geq1$, for which the rotational constraint is lost entirely ($Ro_{\omega}\gg1$ and $Ro_{\text{bulk}}>1$). From Figure \ref{fig:param_summary}d, it is clear that $Ro_{\omega}$ is not diffusion-free (i.e., there is a spread in $Ro_{\omega}$ for a given $Ro_{\text{cv,F}}$). This is because $Ro_{\omega}$ is sensitive to the dissipation scale, which will necessarily be influenced by the diffusivity. However, the dependence of $Ro_{\omega}$ on the diffusivity should not be taken to indicate that the transition between the rotationally-constrained and rotationally-influenced regimes is necessarily diffusivity-dependent; our results will show that these regimes are most accurately demarcated by  $Ro_{\text{cv,F}}$ (which is diffusion-free). Instead, it suggests that $Ro_{\omega}$ (and $Ro_{l}$) is an imperfect measure of the influence of rotation on smaller scales within the inertial range, which, in general, may or may not be sensitive to the diffusivity for a given, intermediate $\ell$.

\begin{figure*}[ht!]
\centering\includegraphics[width=\textwidth]{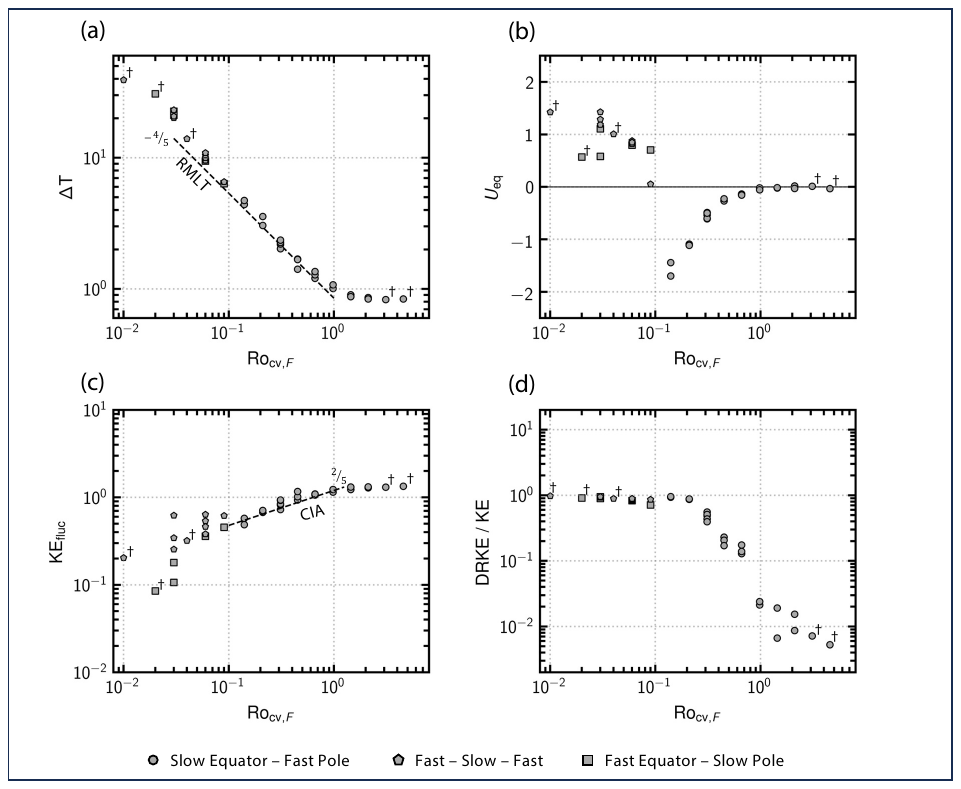}
\caption{Pseudodimensional output data from each experiment. Panel a) shows the shell-averaged temperature difference across the CZ. Panel b shows the equatorial zonal velocity, averaged between $\pm5^{\circ}$ latitude and over the depth of the CZ. Panels c) and d) show the kinetic energy, averaged over the volume of the CZ. KE$_{\text{fluc}}$ is computed using the fluctuating component of the velocity, and DRKE is computed using the axially-symmetric component of the azimuthal velocity. In panels a) and c), lines corresponding to scaling predictions from RMLT and due to CIA balance (see text), respectively, are shown as eye guides. See footnote 4 for details of the re-dimensionalisation procedure. Instances for which only one experiment exists for a given $Ro_{\text{cv,F}}$ are marked with the $\dagger$ symbol. \label{fig:dim_scatter}}
\end{figure*}

All simulations were integrated until the volume averaged kinetic energy and convective heat transport reached a steady state. In practise, this means that each simulation was run for multiple thermal diffusion times (corresponding to $10^{3}$\,--\,$10^{4}$ free-fall times). To ensure numerical convergence, the simulations have been inspected to verify closure of the horizontally-averaged heat budget. The kinetic energy spectra have also been inspected to confirm there is no significant accumulation of energy at the smallest scales so that they are spatially resolved.

\section{Results}\label{sec:results}

\begin{figure*}[t!]
\centering\includegraphics[width=0.925\textwidth]{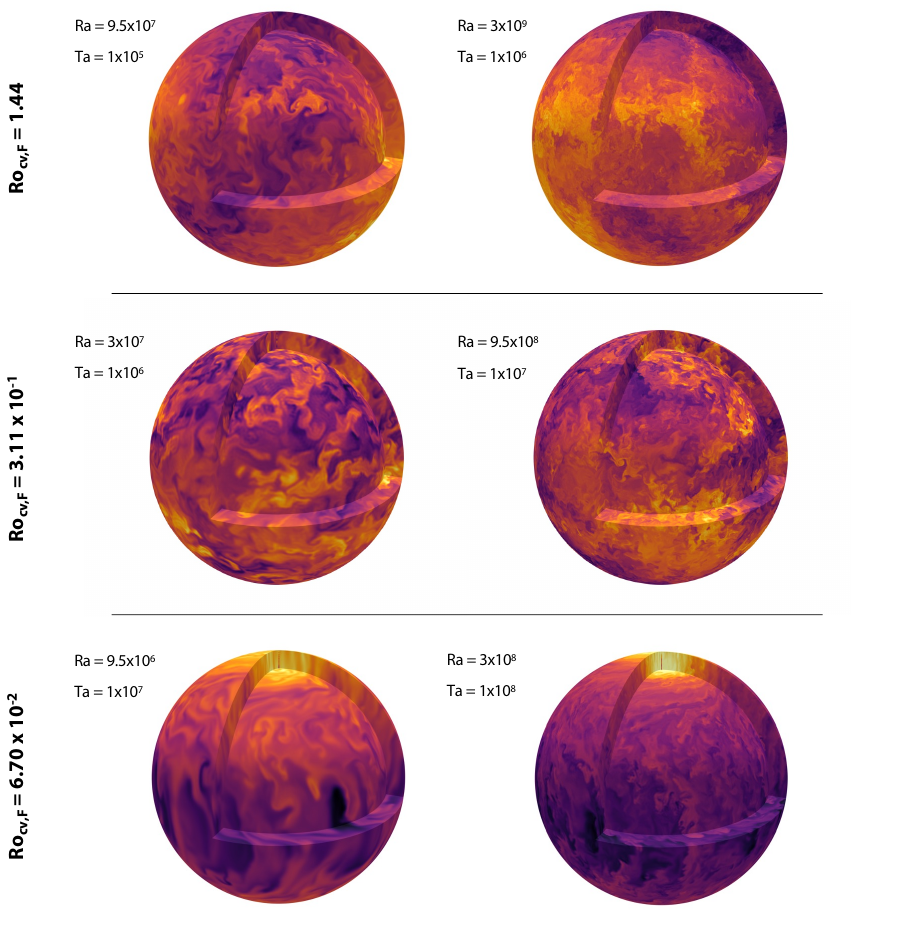}
\caption{Snapshots of the perturbation temperature for six experiments. The experiments shown in the top row are in the rotationally-uninfluenced regime, those in the middle row are in the rotationally-influenced regime, and those in the bottom row are in the rotationally-constrained regime. In each panel, the inner surface corresponds to $r=4.2$ and the outer surface corresponds to $r=4.8$. \label{fig:snaps}}
\end{figure*}

\begin{figure}[t!]
\centering\includegraphics[width=0.4\textwidth]{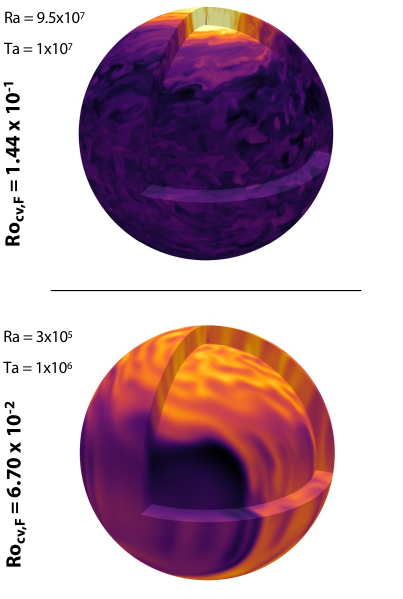}
\caption{Snapshots of the perturbation temperature for two further experiments. The upper panel shows a simulations with $Ro_{\text{cv,F}}=1.44\times10^{-1}$ that is classified as rotationally-influenced. In this experiment, the influence on the dynamics is more apparent visually than is the case for the two rotationally-influenced simulations with $Ro_{\text{cv,F}}=3.11\times10^{-1}$ shown in Figure \ref{fig:snaps}. The lower panel shows a rotationally constrained case with $Ro_{\text{cv,F}}=6.70\times10^{-2}$ (as in Figure \ref{fig:snaps}, bottom row). This simulation is more dissipative than those with $Ro_{\text{cv,F}}=6.70\times10^{-2}$ shown in Figure \ref{fig:snaps}. In the upper panel, the colourbar maximum has been rescaled by $0.7$ to enhance the visibility of features at lower latitudes. \label{fig:extra_snaps}}
\end{figure}

\subsection{Simulation results}

Summary statistics for our experiments are presented in Figure \ref{fig:dim_scatter}. All quantities shown in this figure are pseudodimensional (specifically, they have been re-dimensionalised using $F=1$ and $d=1$ to set the units\footnote{Taking $F=1$ and $d=1$, dimensional variables are obtained from their non-dimensional counterparts using $T_{\text{dim}}=\Lambda T$ and $\boldsymbol{u}_{\text{dim}}=\boldsymbol{u}/\tau$, where $\Lambda=1/\kappa$ and $\tau=\kappa^{\frac{1}{2}}$ (Section \ref{sec:nd}). $\kappa$ is obtained from the definition of $Ra_{\text{F}}$ (using $Pr=1$).}), and are plotted against $Ro_{\text{cv,F}}$ (equivalent to $1/(2\Omega_{0})$ for our choice of re-dimensionalisation; cf. Equation \ref{eq:Rocv}). In these panels, diffusivity-dependence manifests as scatter on the $y$-axis for a given value $Ro_{\text{cv,F}}$ (note: not all $Ro_{\text{cv,F}}$ have multiple simulations; those with only one entry are marked with a $\dagger$ symbol). Conversely, if the dynamics are diffusion-free, then all the data for a given $Ro_{\text{cv,F}}$ will collapse onto one point.

Figure \ref{fig:dim_scatter}a shows the shell-averaged temperature difference $\Delta T$ between the inner and outer radii of the CZ. For $Ro_{\text{cv,F}}<1$, $\Delta T$ displays a strong dependence on $Ro_{\text{cv,F}}$, increasing with $\Omega_{0}$ as the rotational constraint acts to throttle convective heat transport. For each $Ro_{\text{cv,F}}$, the scatter between experiments with different $Ra_{\text{F}}$ is very small. In the rotationally-influenced regime, the dependence of $\Delta T$ on $Ro_{\text{cv,F}}$ ($\Delta T\propto\Omega_{0}^{\frac{4}{5}}\,{\sim}\,Ro_{\text{cv,F}}^{-\frac{4}{5}}$) is consistent with rotating mixing length theory (RMLT; see Section \ref{sec:theory}; \citealp{1979GApFD..12..139S,2014ApJ...791...13B}), which is a diffusion-free theory. For the cases categorised as rotationally-constrained with $Ro_{\text{cv,F}}<0.1$, $\Delta T$ depends more strongly on $Ro_{\text{cv,F}}$ than predicted by RMLT, while remaining independent of the diffusivity (for the values of $Ro_{\text{cv,F}}$ for which there are multiple experiments). In the rotationally-uninfluenced regime where $Ro_{\text{cv,F}}>1$, the rotational constraint is lost, causing $\Delta T$ to become independent of $Ro_{\text{cv,F}}$. Our simulations yield results that are consistent with this expectation, and thus by extension are consistent with diffusion-free behaviour. 

The azimuthal velocity averaged within $\pm5^{\circ}$ of the equator and throughout the depth of the CZ, $U_{\text{eq}}$, is shown in Figure \ref{fig:dim_scatter}b. The dependence of $U_{\text{eq}}$ on $Ro_{\text{cv,F}}$ we obtain is consistent with previous work \citep{2014MNRAS.438L..76G}. In particular, the transition between prograde and retrograde equatorial flow occurs when $Ro_{\text{cv,F}}\approx0.1$, which corresponds to $Ro_{\omega}\approx1$ (and $Ro_{l}\approx1$; not shown). In the absence of a rotational constraint ($Ro_{\text{cv,F}}>1$), differential rotation does not develop ($U_{\text{eq}}\approx0$). In the rotationally-influenced regime ($Ro_{\text{cv,F}}<1$; $Ro_{\omega}\gtrsim1$), anti-Solar (retrograde) differential rotation develops, with an amplitude that increases as the rotation rate is increased. In this regime, the amplitude of the equatorial zonal flow is independent of the diffusivity. When rotation becomes dominant ($Ro_{\text{cv,F}}<0.1$; $Ro_{\omega}\lesssim1$), the differential rotation switches direction from retrograde to prograde (Solar-like). In this regime, the data in general do not collapse onto a single curve for $U_{\text{eq}}$ vs. $Ro_{\text{cv,F}}$ indicating that, at least for the supercriticalities considered in the present work, the strength of $U_{\text{eq}}$ depends on the diffusivities. We note that there are only two values of $Ro_{\text{cv,F}}<0.1$ for which we ran a significant number of experiments with varying $Ra_{\text{F}}$ (5 experiments each for $Ro_{\text{cv,F}}=3.11\times10^{-2}$ and $6.70\times10^{-2}$). For the series with $Ro_{\text{cv,F}}=6.70\times10^{-2}$, $U_{\text{eq}}$ does not vary significantly as the diffusivity is varied, but for the series with $Ro_{\text{cv,F}}=3.11\times10^{-2}$ there is a diffusivity dependence. 

Figures \ref{fig:dim_scatter}c and d show the kinetic energy (KE) averaged over the volume of the CZ. Figure \ref{fig:dim_scatter}c shows $\text{KE}_{\text{fluc}}$, computed using the r.m.s. fluctuating velocity, and Figure \ref{fig:dim_scatter}d shows the ratio DRKE/KE, where KE is the total kinetic energy, and DRKE is the `differential rotation kinetic energy', computed using the axially-symmetric component of the azimuthal velocity. The behaviour of $\text{KE}_{\text{fluc}}$ in Figure \ref{fig:dim_scatter}c tells a similar story to $U_{\text{eq}}$, namely that the kinematic properties of the flow display a weak dependence on the diffusivity for $Ro_{\text{cv,F}}>0.1$ (rotationally-uninfluenced or rotationally-influenced), but a stronger dependence once the influence of rotation becomes dominant. There are two additional features of Figure \ref{fig:dim_scatter}c,d that are worth emphasising. First, we note that in the rotationally-influenced regime, the dependence of $\text{KE}_{\text{fluc}}$ on $Ro_{\text{cv,F}}$ is consistent with the diffusion-free `Coriolis-inertial-Archimedean' (CIA) scaling (e.g., \citealp{2016JFM...808..690G}; see Section \ref{sec:theory}). By contrast, in the rotationally-constrained regime, $\text{KE}_{\text{fluc}}$ is diffusivity-dependent. However, the ratio DRKE/KE remains roughly diffusion-free. This is in large part due to the fact that the DRKE dominates the total KE in this regime, so that their ratio saturates at $\text{DRKE}/\text{KE}\approx1$, although we note there is some compensation between the diffusivity dependence of the DRKE and $\text{KE}_{\text{fluc}}$ (i.e., for a given $Ro_{\text{cv,F}}$ an experiment with a larger $\text{KE}_{\text{fluc}}$ tends to have a larger DRKE).


The dependence of the summary statistics shown in Figure \ref{fig:dim_scatter} on $Ro_{\text{cv,F}}$ (and the diffusivities, where applicable) arise from the influence of these parameters on the convective flow patterns. To illustrate this, Figure \ref{fig:snaps} shows the perturbation temperature $T^{\prime}$ (horizontal average removed) for six selected experiments. The upper row shows two experiments with $Ro_{\text{cv,F}}=1.44$ that occupy the rotationally-uninfluenced regime, the middle row shows two experiments with $Ro_{\text{cv,F}}=3.11\times10^{-1}$ that fall within the rotationally-influenced regime, and finally the bottom row shows two experiments with $Ro_{\text{cv,F}}=6.70\times10^{-2}$ that occupy the rotationally-constrained regime. 

In the rotationally-uninfluenced regime, convective plumes extend outwards in the radial direction at all latitudes and the effect of rotation is undetectable. In this regime, the temperature field is organised into large `warm' and `cool' patches that are dominated by spherical harmonic degree $\ell=1$ or $2$. The morphology in the rotationally-influenced cases with $Ro_{\text{cv,F}}=3.11\times10^{-1}$ is similar, although the convection at higher latitudes now exhibits a weak alignment with the rotation axis and some low-latitude features are elongated in the azimuthal direction. However, the impact of rotation is hard to detect, consistent with $Ro_{\omega}>1$. Figure \ref{fig:extra_snaps} shows an additional case in the rotationally-influenced regime with $Ro_{\text{cv,F}}=1.44\times10^{-1}$ (i.e., a slightly faster rotation rate), which more clearly shows that the convection is aligned cylindrically, and zonally-elongated at lower latitudes. We highlight that the pair of experiments with $Ro_{\text{cv,F}}=3.11\times10^{-1}$ are morphologically very similar, as are the pair with $Ro_{\text{cv,F}}=1.44$, which is consistent with the inference from Figure \ref{fig:dim_scatter} that the diffusivity has a limited influence on the dynamics in the rotationally-uninfluenced and rotationally-influenced regimes. 

The effect of rotation in the rotationally-constrained regime is clear. In mid-latitudes and the polar regions, convection is aligned cylindrically (i.e., with the rotation axis). Closer to the equator, large, modulated thermal Rossby wave-like features can be identified, that propagate in the prograde direction (with respect to the mean zonal flow). These waves are present in all experiments with $Ro_{\text{cv,F}}<0.1$ but in none with $Ro_{\text{cv,F}}\geq1$. The transition to a regime of equatorial dynamics dominated by wave-like features occurs when small and intermediate spatial scales become rotationally constrained ($Ro_{\omega}<1$). For the relatively turbulent cases shown in Figure \ref{fig:snaps} there is a strong polar vortex. This is not present in simulations that are more strongly dissipative (an example is given by the lower panel of Figure \ref{fig:extra_snaps}).

\begin{figure*}[ht!]
\centering\includegraphics[width=\textwidth]{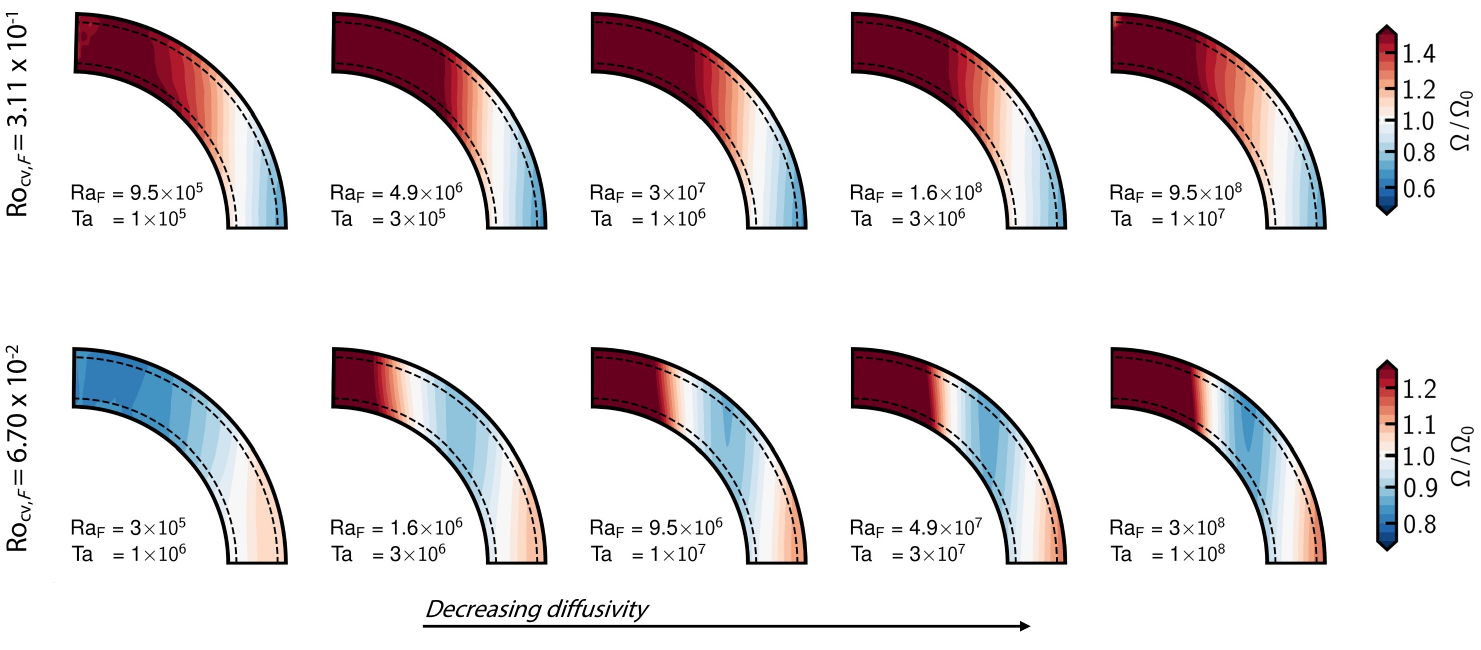}
\caption{Azimuthally-averaged  angular velocity $\Omega$ normalised by the rotation rate $\Omega_{0}$ (see Equation \ref{eq:Om}) for two series of simulations with fixed $Ro_{\text{cv,F}}$. The top row shows $Ro_{\text{cv,F}}=3.11\times10^{-1}$ and the bottom row shows $Ro_{\text{cv,F}}=6.70\times10^{-2}$. For a given $Ro_{\text{cv,F}}$, different combinations of $Ra_{\text{F}}$ and $Ta$ correspond to different diffusivities. In the slices above, the diffusivity decreases from left to right. The dashed lines show the edges of the CZ at $r_{\text{i}}=4$ and $r_{\text{o}}=5$. \label{fig:DR}}
\end{figure*}

Figure \ref{fig:DR} shows cross-sections of the differential rotation, quantified by the azimuthally-averaged angular velocity normalised by the rotation rate $\Omega_{0}$, \begin{equation} 
\frac{\Omega}{\Omega_{0}} = 1 + \frac{\overline{u}_{\phi}}{\Omega_{0} r\sin\theta}, \label{eq:Om}
\end{equation} 
for two series of experiments at $Ro_{\text{cv,F}}=3.11\times10^{-1}$ (rotationally-influenced; shown in the upper row) and $Ro_{\text{cv,F}}=6.70\times10^{-2}$ (rotationally-constrained; lower row). For a given $Ro_{\text{cv,F}}$, larger $Ra_{\text{F}}$ corresponds explicitly to reduced diffusivity (decreasing across panels moving from left to right). The experiments in the $Ro_{\text{cv,F}}=3.11\times10^{-1}$ set exhibit `anti-Solar' differential rotation with a slow equator and a fast pole (denoted SF). The differential rotation in this set of experiments has little dependence on the diffusivity, with the spatial structure and amplitude of $\Omega/\Omega_{0}$ remaining essentially unchanged as $Ra_{\text{F}}$ is increased. In contrast, the differential rotation in the $Ro_{\text{cv,F}}=6.70\times10^{-2}$ set does depend on the diffusivity. All of these experiments have a fast equator, but the structure of the zonal flow at higher latitudes varies. The most diffusive experiment (left-most panel in Figure \ref{fig:DR}, bottom row) features a Solar-like fast equator and slow pole (denoted FS), while the remaining four experiments have prograde flow at the equator, flanked by retrograde flow in mid-latitudes, and then finally a polar vortex (cf. Figure \ref{fig:snaps}d) at high-latitudes (i.e., fast--slow--fast; denoted FSF). Once the transition from FS to FSF differential rotation occurs, the $Ro_{\text{cv,F}}=6.70\times10^{-2}$ experiments appear to converge towards an asymptotic `diffusion-free' structure and amplitude as the diffusivity is reduced further (see also the small spread in $U_{\text{eq}}$ when $Ro_{\text{cv,F}}=6.70\times10^{-2}$ in Figure \ref{fig:dim_scatter}).

 Each of the scatter charts in this article use the marker shape to illustrate the occurrence of the SF, FSF, and FS configurations of differential rotation in parameter space (circles for SF, pentagons for FSF, and squares for FS). From Figure \ref{fig:param_summary}a--c, we identify that solar-like differential rotation (FS) is only obtained for the most diffusive experiments ($Ra_{\text{F}}\lesssim10^{6}$), while more turbulent simulations with a strong rotational constraint exhibit FSF differential rotation. Figure \ref{fig:param_summary}d shows that the transition from a fast equator (either FS or FSF) to a slow equator occurs at the boundary between the rotationally-constrained and rotationally-influenced regimes, when $Ro_{\text{cv,F}}=0.1$, which corresponds to $Ro_{\omega}\approx1$, consistent with previous work \citep{2014MNRAS.438L..76G}. The fact that all of the FS or FSF cases have $Ro_{\text{cv,F}}<0.1$, and all of the SF cases have $Ro_{\text{cv,F}}\geq0.1$, implies that the transition from a fast equator to a slow equator is diffusion-free (even if the zonal flow structure in the fast-equator cases themselves displays a dependence on the diffusivity). We believe that this transition is associated with the existence of thermal Rossby wave-like features at low latitudes, which are present in all simulations in the rotationally-constrained regime but absent otherwise. However, a complete analysis of the zonal momentum budget is left for future work. 
 
Finally, we note that the experiments we categorise as rotationally-constrained were run with a lower supercriticality than those categorised as rotationally-influenced or rotationally-uninfluenced (Figure \ref{fig:param_summary}b). While we do not observe it here, we would expect the anti-Solar differential rotation in these regimes to display a diffusivity-dependence for sufficiently low supercriticality (i.e., higher diffusivity).

\begin{figure*}[t!]
\centering\includegraphics[width=\textwidth]{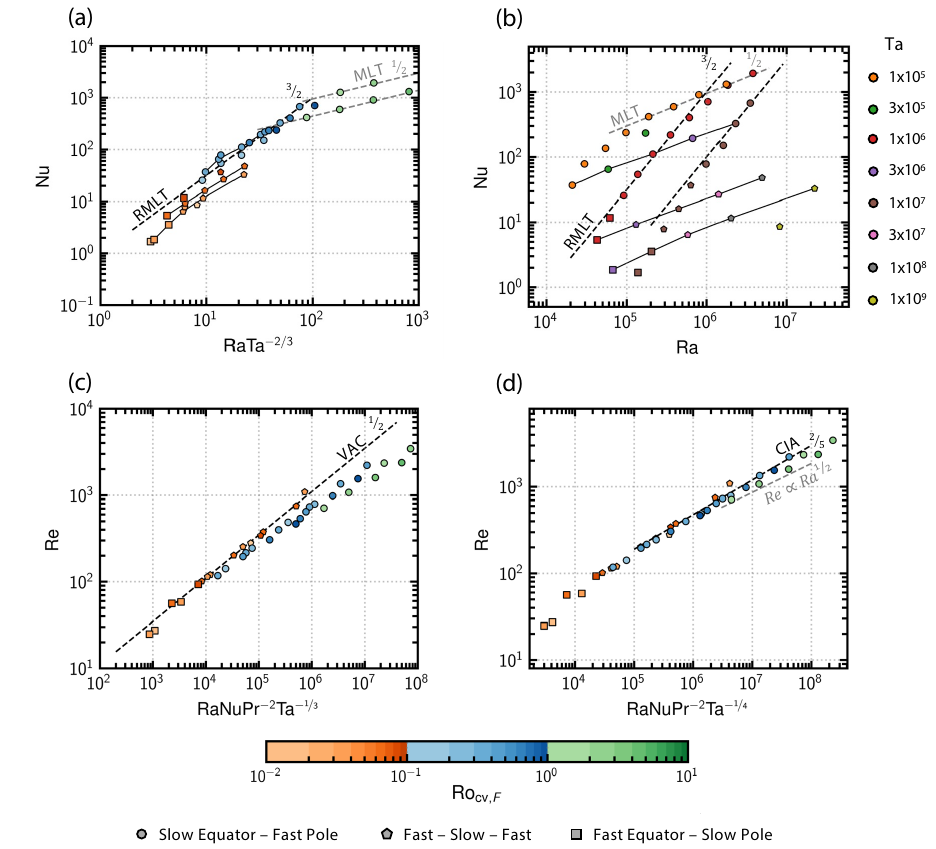}
\caption{Scaling results for $Nu$ (panels a and b) and $Re$ (panels c and d) obtained with our experiments. In panels a) and b), grey and black dashed lines correspond to the mixing length theory (MLT) and rotating mixing length theory (RMLT) scaling laws given by Equations \ref{eq:NuMLT} and \ref{eq:NuRMLT}, respectively, and the black solid lines connect selected series of experiments with constant $Ro_{\text{cv,F}}$ (see caption of Figure \ref{fig:param_summary} a--c for further information). In panel c), the scaling for $Re$ predicted by VAC balance (Equation \ref{eq:ReVAC}) is shown as a black dashed line. Finally, in panel d), the black dashed line shows the inertial scaling of rotating convection (associated with CIA balance; given by Equation \ref{eq:ReCIA}), and the grey dashed line shows the diffusion-free, non-rotating `ultimate' scaling $Re\propto Ra^{1/2}$ (Equation \ref{eq:ReMLT}).}\label{fig:nondim_scatter}
\end{figure*}

\subsection{Comparison with scaling predictions}\label{sec:theory}

\subsubsection{Nusselt number scaling}

Radial heat transport is often characterized using the Nusselt Number, defined as the ratio of the shell-integrated total heat flux (convective plus conductive) to the conductive flux. In this work, we evaluate this as an integral over the CZ, which yields: \begin{equation} 
Nu\equiv\frac{\mathscr{F}_{\text{tot}}}{\mathscr{F}_{\text{cond}}}=\frac{r_{\text{i}}^{2}}{\int_{r_i}^{r_o}\partial_{r}\langle T\rangle\,r^{2}\text{d}r}. \label{eq:Nu}
\end{equation}
All quantities above are non-dimensional, $\langle T\rangle$ is the shell-averaged temperature, and the second relation follows from the fact that $\mathscr{F}_{\text{tot}} = 4\pi r_{\text{i}}^{2}$ within the CZ.  Mixing length theory, which essentially equates the kinetic energy of fluid parcels to the buoyancy work done over a characteristic length $l$ (the mixing length), states that the Nusselt number should scale as: \begin{equation} 
Nu\propto Ra^{\frac{1}{2}}, \label{eq:NuMLT}
\end{equation}
for $Pr=1$. A central assumption in MLT is that the total dimensional heat transport is independent of the diffusivities, and thus the scaling above is diffusion-free. We note that a simple dimensional analysis for a non-rotating system yields the scaling above under this assumption\footnote{A diffusion-free scaling for a non-dimensional number (e.g., $Nu$ or $Re$) does not mean that the scaling parameter (e.g., $Ra$) is diffusivity-independent. Instead, it is a scaling that is consistent with the dimensional quantities (e.g., the dimensional radial temperature gradient for $Nu$) being diffusivity-independent. For the specific case of the MLT $Nu$ scaling, $Nu$ includes a factor of $\kappa^{-1}$ if $T$ is re-dimensionalised, and so the right-hand-side must also contain a factor of $\kappa^{-1}$, which implies a scaling $Nu\propto Ra^{\frac{1}{2}}Pr^{\frac{1}{2}}$ (which is $Ra^{\frac{1}{2}}$ for $Pr=1$).}. For a rotating system, the analogous theory to MLT is rotating mixing length theory (RMLT; \citealp{1979GApFD..12..139S}), which assumes that convection is dominated by modes with a growth rate well described by linear theory and amplitudes determined by a balance between their growth rate and the non-linear cascade rate. RMLT predicts that $Nu$ scales as  \begin{equation} 
Nu\propto\frac{Ra^{\frac{3}{2}}}{TaPr^{\frac{1}{2}}}. \label{eq:NuRMLT}
\end{equation}
As with Equation \ref{eq:NuMLT}, this is a diffusion-free scaling. This scaling can also be obtained by combining the `diffusion-free heat transport assumption' with a further assumption that the heat transport depends only on the supercriticality of the system $Ra/Ra_{\text{c}}$ (recall that $Ra_{\text{c}}\,{\sim}\,Ta^{\frac{3}{2}}$ for large $Ta$) \citep{2012PhRvL.109y4503J, 2014ApJ...791...13B}. We note that $Ra\approx Ra_{\text{F}}/Nu$, which yields the scaling $Nu\propto\left(Ra_{\text{F}}Ta^{-2/3}\right)^{3/5} Pr^{-1/5}$. 
Using the definition of $Ro_{\text{cv,F}}$, this can be re-written 
$Nu\propto Ra_{\text{F}}^{1/3}Pr^{1/3}Ro_{\text{cv,F}}^{4/5}$. Therefore, with $Pr$ constant, RMLT implies \begin{equation}
Nu\propto Ra^{\frac{1}{2}}\ \ \text{for constant}\ Ro_{\text{cv,F}} \label{eq:NuRMLTRoCV}
\end{equation}
(i.e., as in MLT). Previous studies of boundary-driven Boussinesq convection in a spherical shell have not yielded MLT-like scaling for the Nusselt number in the regime where rotation is unimportant. Instead, shallower scaling exponents were obtained, due to the throttling of convection by the conductive heat transport in the boundary layers (e.g., \citealp{2015JFM...778..721G}). Some studies have obtained RMLT-like scaling in the rapidly rotating regime, but only for extreme parameter values (e.g., \citealp{2016JFM...808..690G}, wherein experiments with $Ta\gtrsim10^{12}$ and $RaTa^{-2/3}\simeq10$ exhibit a Nusselt number scaling in agreement with RMLT).

Figure \ref{fig:nondim_scatter}a and \ref{fig:nondim_scatter}b show $Nu$ plotted against $RaTa^{-2/3}$ in panel a) and $Ra$ in panel b). Dashed lines corresponding to the MLT and RMLT scaling predictions are included in each panel (note that the MLT scaling in panel a, and the RMLT scaling in panel b, apply to series of experiments with constant $Ta$). In panel a), the marker colour is used to show $Ro_{\text{cv,F}}$. 

Inspection of Figure \ref{fig:nondim_scatter}a reveals that our experiments fall into three groups distinguished by different $Nu$ vs. $RaTa^{-2/3}$ scaling behaviours. When $Ro_{\text{cv,F}}\geq1$ (indicated by a green marker colour), the Nusselt number follows the MLT scaling prediction (Equation \ref{eq:NuMLT}). These experiments are those that we have categorised as rotationally-uninfluenced. For, $0.1\leq Ro_{\text{cv,F}}<1$ (indicated by blue marker colours), i.e., experiments in the rotationally-influenced regime, a steeper scaling is obtained that is consistent with RMLT (Equation \ref{eq:NuRMLT}). The transition between the MLT and RMLT scalings occurs when $Ro_{\text{cv,F}}=1$, which roughly corresponds to the point where $Ro_{\text{bulk}}=1$. The agreement between the scaling obtained by the rotationally-influenced experiments and RMLT is striking, given that the effect of rotation on the convective morphology in this regime can be relatively weak (e.g., Figure \ref{fig:snaps}). These results (as well as those presented in the next subsection) inform our definition of the rotationally-uninfluenced and rotationally-influenced regimes. Finally, in the rotationally-constrained regime for which $Ro_{\text{cv,F}}<0.1$ (orange marker colours), the Nusselt number scaling departs from the RMLT branch occupied by the rotationally-influenced cases. This departure from RMLT is manifest as an offset  from the RMLT branch (reduced $Nu$ for a given $RaTa^{-2/3}$) that increases as $Ro_{\text{cv,F}}$ is reduced. In Figure \ref{fig:nondim_scatter}b, it is apparent that for constant $Ro_{\text{cv,F}}$, experiments in this regime roughly follow the $Nu\propto Ra^{1/2}$ scaling noted above (Equation \ref{eq:NuRMLTRoCV}; for simulations with sufficient supercriticality, $RaTa^{-2/3}\gtrsim5$). This indicates that the convective heat transport is still independent of the diffusivity (consistent with the pseudodimensional $\Delta T$ shown in Figure \ref{fig:dim_scatter}a), and that the offset from the rotationally-influenced RMLT branch captures a dependence of the convective heat transport on $Ro_{\text{cv,F}}$ (a diffusion-free parameter) that is not accounted for by RMLT. Hereafter, we refer to this scaling as `offset RMLT'.  As noted previously, the transition from the rotationally-influenced regime to this new regime occurs when $Ro_{\omega}\lesssim1$, which indicates that all spatial scales are now affected by rotation, motivating our choice to refer to this regime as rotationally-constrained.

When the same data are plotted against $Ra$ (Figure \ref{fig:nondim_scatter}b), the MLT scaling obtained by the rotationally-uninfluenced regime can be identified for experiments with $Ta=10^{5}$, which unambiguously follow the $Ra^{1/2}$ prediction. The $Ta=10^{6}$ series follows the RMLT scaling for moderate supercriticality, before transitioning towards the MLT scaling at higher supercriticality, and the experiments at $Ta=10^{7}$ are approaching the RMLT scaling. Finally, the offset RMLT scaling identified for the rotationally-constrained regime is manifest as a steepening of the $Nu(Ra)$ relation for fixed $Ta$ (but, again, note that series of experiments with constant $Ro_{\text{cv,F}}$, joined by black solid lines, follow the scaling predicted by RMLT, in the form given by Equation \ref{eq:NuRMLTRoCV}).

A preliminary analysis (not shown) of the latitudinal dependence of the heat budget and Nusselt number (cf. \citealp{2023JFM...954R...1G}) suggests that the transition to the offset RMLT scaling that characterises the rotationally-constrained regime may be related to the change in the dynamics from a flow dominated by turbulent convection to a wave-dominated flow (we find wave-like features are always present when $Ro_{\text{cv,F}}<0.1$, but are absent when $Ro_{\text{cv,F}}\geq0.1$). However, the offset scaling could alternatively arise due to interaction between the convection and the zonal flow, although this would have to be a `signed' effect, since strong (anti-Solar) zonal flows are also present in the rotationally-influenced regime for which the standard RMLT scaling is recovered. We will explore this further in future work. We note that the offset scaling persists even for the rotationally-constrained experiments with the highest supercriticality. As a result, we do not believe that it arises because the rotationally-constrained simulations generally have lower supercriticality than those categorised as rotationally-influenced.

\addtolength{\tabcolsep}{-0.2em}
\begin{deluxetable*}{l|ccc|cc|cc|c}
\tablewidth{0pt}
\tablecaption{Summary of dynamical regimes. \label{tab:1}}
\tablehead{
\colhead{} & \multicolumn{3}{c}{$Ro$} & \multicolumn{2}{c}{$Nu$ / $\Delta T$} & \multicolumn{2}{c}{$Re$ / KE$_{\text{fluc}}$} & \colhead{} \\[-4pt] 
\colhead{Name} & \colhead{$Ro_{\text{cv,F}}$} & \colhead{$Ro_{\omega}$}& \colhead{$Ro_{\text{bulk}}$} &\colhead{Diff.-free?} & \colhead{Scaling} & \colhead{Diff.-free?} & \colhead{Scaling} & \colhead{Differential Rotation}
}
\startdata
Rot.-constrained & $Ro<0.1$ & $Ro\lesssim1$  & $Ro\ll1$ & \ding{51} & Offset RMLT & \ding{55} & VAC & FS or FSF\\
Rot.-influenced & $0.1\leq Ro<1$ & $Ro\gtrsim1$ & $Ro<1$ & \ding{51} & RMLT & \ding{51} & CIA & SF \\
Rot.-uninfluenced & $Ro\geq1$ & $Ro\gg1$ & $Ro\geq1$ & \ding{51} & MLT & \ding{51} & Ultimate & SF
\enddata
\tablecomments{Experiments are classified according to their input $Ro_{\text{cv,F}}$. Two output Rossby numbers, $Ro_{\omega}$ and $Ro_{\text{bulk}}$, are also included in the table. These are defined in Equation \ref{eq:Ro_alt}, and are intended to indicate the degree of rotational constraint at small scales, and the largest spatial scale, respectively. $Nu$ is the Nusselt Number (Equation \ref{eq:Nu}), and $\Delta T$ is the radial temperature contrast. Both are evaluated for the convection zone (CZ; between $r_{\text{i}}=4$ and $r_{\text{o}}=5$). `MLT' and `RMLT' refer to `mixing length theory' and `rotating mixing length theory', respectively. The scalings for $Nu$ associated with these theories are given by Equations \ref{eq:NuMLT} and \ref{eq:NuRMLT}. $Re$ is the Reynolds number (Equation \ref{eq:Re}), and $\text{KE}_{\text{fluc}}$ is the fluctuating kinetic energy. Both are computed using the r.m.s. fluctuating (azimuthal average removed) velocity within the CZ. `VAC' and `CIA' refer to `Viscous-Archimedean-Coriolis balance' and `Coriolis-inertial-Archimedean balance', respectively. The VAC, CIA, and `Ultimate' scalings are given by Equations \ref{eq:ReMLT}--\ref{eq:ReVAC}. Finally, the differential rotation classifications `FS', `FSF', and 'SF' correspond to configurations with: (i) a fast equator and slow pole (Solar-like), (ii) a fast equator, slow mid-latitudes, and `fast' polar vortex, and (iii) a slow equator and fast pole (anti-Solar). \vspace*{-.3in}}
\end{deluxetable*}

\subsubsection{Reynolds number scaling}

A common metric of the degree of turbulence in fluid flow is the Reynolds number $Re$ (given by Equation \ref{eq:Re}), which quantifies the ratio of the inertial and viscous terms in the momentum equation. In the non-rotating limit, dimensional analysis predicts the `ultimate' scaling \begin{equation}
Re\propto Ra^{\frac{1}{2}}\label{eq:ReMLT}
\end{equation}
under the assumption that the velocity $U$ (used to compute $Re$) and the temperature difference $\Delta T$ (used to compute $Ra$) are independent of the diffusivities. When instead the system is rotating, then a diffusion-free balance between the Coriolis, inertial, and Archimedean (buoyancy driving) terms in the momentum equation (CIA balance) suggests the scaling \citep{2001PEPI..128...51A,2014ApJ...791...13B,2016JFM...808..690G,2021PNAS..11822518V}: \begin{equation} 
Re\propto \left(RaNuPr^{-2}Ta^{-\frac{1}{4}}\right)^{\frac{2}{5}},\label{eq:ReCIA}
\end{equation}
sometimes referred to as the inertial scaling of rotating convection. 
If alternatively, a balance is sought between the Coriolis, viscous, and Archimedean terms (i.e., VAC balance; \citealp{2013JFM...717..449K}), then the following scaling: \begin{equation} 
Re\propto\left(RaNuPr^{-2}Ta^{-\frac{1}{3}}\right)^{\frac{1}{2}}\label{eq:ReVAC}
\end{equation}
is obtained. Equation \ref{eq:ReVAC} implies that $U$ depends on the diffusivity. We are not aware of any study that recovers either of the diffusion-free scalings (Equations \ref{eq:ReMLT} and \ref{eq:ReCIA}) from simulations of spherical shell convection that include rotation. For example, even the simulations presented in \citet{2016JFM...808..690G}, that achieve an RMLT-like scaling for $Nu$ in the rapidly rotating regime, exhibit a Reynolds number scaling that is not diffusion-free. There the rapidly rotating cases follow an alternate scaling in which CIA balance is obtained in the bulk, but kinetic energy dissipation is dominated by the boundary layers, while their weakly-nonlinear simulations follow the VAC scaling described above. However, we note that \citet{2016ApJ...818...32F} do find diffusion-free behaviour for the kinetic energy in simulations that are non-rotating.

Figure \ref{fig:nondim_scatter}c and \ref{fig:nondim_scatter}d present a comparison between the Reynolds number scaling obtained in our simulations, and the theoretical relations given by Equations \ref{eq:ReMLT}--\ref{eq:ReVAC}. As with the Nusselt number, the three regimes we have identified correspond to three distinct scaling relations. Experiments with $Ro_{\text{cv,F}}>1$ in the rotationally-uninfluenced regime, for which all scales are rotationally-unconstrained, follow the ultimate scaling given by Equation \ref{eq:ReMLT} (green points in Figure \ref{fig:nondim_scatter}d). Meanwhile, the experiments that are rotationally-influenced ($Ro_{\text{bulk}}<1$ but $Ro_{\omega}\gtrsim1$) follow the inertial scaling (blue points in Figure \ref{fig:nondim_scatter}d). By contrast, the rotationally-constrained $Ro_{\text{cv,F}}<0.1$ cases display a Reynolds number scaling that closely follows the diffusivity-dependent VAC balance scaling (orange points in Figure \ref{fig:nondim_scatter}c), which is consistent with the dependence of the KE on the diffusivities in Figure \ref{fig:dim_scatter}.

\section{Summary and Discussion}\label{sec:discuss}

We have presented 3D simulations of rotating convection in a spherical shell, driven by an internal heating and cooling function.  This function was designed so that the same flux is extracted by cooling (over a fixed region near the top of the computational domain) as is input by heating (over an identically-sized region near the base).  A key finding of our work is that in some regimes, the heat transport and other aspects of the flow dynamics are well described by diffusion-free theory, with no evident dependence on the viscosity or thermal diffusivity. Below, we briefly summarise the main results of this study, relate them to prior work, and comment on their astrophysical implications.

The scaling results presented in the previous section suggest that our simulations can be grouped into three distinct regimes, which we term rotationally-constrained ($Ro_{\text{cv,F}}<0.1$), rotationally-influenced ($0.1\leq Ro_{\text{cv,F}}<1$), and rotationally-uninfluenced ($Ro_{\text{cv,F}}\geq1$). In the rotationally-constrained regime, all spatial scales are constrained by rotation. The rotationally-influenced regime is an intermediate regime where largest scales are constrained by rotation, but some intermediate scales are not (Figure \ref{fig:param_summary}d). Finally, there is no discernible impact of rotation on the dynamics of simulations in the rotationally-uninfluenced regime.  The key features of each regime are summarised in Table \ref{tab:1} and described below.  

 The main result of this paper is that all of our experiments (aside from those that are very weakly supercritical) exhibit Nusselt number scalings and a radial temperature contrast that are independent of the diffusivities (Figure \ref{fig:dim_scatter}a, Figure \ref{fig:nondim_scatter}a,b). Simulations in the non-rotating and rotationally-influenced regimes follow scalings that are consistent with MLT and RMLT, respectively; the rotationally-constrained cases also follow a scaling that is RMLT-like, but offset by a factor of the (diffusion-free) convective Rossby number ($Ro_{\text{cv,F}}$). In addition, diffusion-free scaling is also obtained for the Reynolds number in the non-rotating and rotationally-influenced regimes (Figure \ref{fig:nondim_scatter}d), but in the rotationally-constrained regime the experiments follow a Reynolds number scaling consistent with VAC balance (which depends on the diffusivity; Figure \ref{fig:nondim_scatter}c). Finally, in the rotationally-constrained regime, the morphology of the differential rotation depends on the supercriticality; weakly supercritical simulations ($RaTa^{-2/3}\lesssim4$) feature Solar-like differential rotation (denoted FS), whereas moderately supercritical simulations additionally feature a strong, prograde polar vortex (denoted FSF) (as in Figure \ref{fig:DR}, bottom row). In the rotationally-influenced and non-rotating regimes, anti-Solar differential rotation (denoted SF) is obtained (as in Figure \ref{fig:DR}, top row), with a structure and amplitude that is independent of the diffusivity.

An important result from our simulations is that it is possible to obtain zonally-averaged zonal flow statistics that are independent of the diffusivity. This result is demonstrated by rotationally-influenced cases with finite amplitude anti-Solar differential rotation (Figure \ref{fig:dim_scatter}b and Figure \ref{fig:DR}). In the rotationally-constrained regime, the differential rotation does depend on the diffusivity, although we suggest that too few of our experiments have sufficient supercriticality (only six experiments with $RaTa^{-2/3}>10$ and $Ro_{\text{cv,F}}<0.1$) to explore the possibility for diffusion-free prograde differential rotation. Notwithstanding the diffusivity-dependence of the differential rotation in the rotationally-constrained regime, we have found that the sign of the equatorial zonal flow (i.e., prograde -- FS or FSF, or retrograde -- SF), does not depend on the diffusivity.

Previous work has shown that `local Nusselt numbers' defined for the equatorial and polar regions can yield different scaling results that are set by the local dynamics \citep{2023JFM...954R...1G}. Therefore, the fact that RMLT does not fully capture the convective heat transport scaling in the rotationally-constrained experiments is not surprising, given that the morphology of the convection depends strongly on latitude (e.g., wave-like dynamics at low latitudes but convective turbulence at high latitudes) but this effect is not accounted for by RMLT. A full analysis of the latitudinal heat budget, as well as the dependence of the size of the equatorial region on $Ro_{\text{cv,F}}$, is required to form a complete understanding of the $Nu$ scaling in the rotationally-constrained regime. A second, puzzling feature of the rotationally-constrained regime is that the convective velocity scaling (i.e., the $Re$ scaling) is diffusivity-dependent, but the heat transport scaling is not. This discrepancy may simply imply that the r.m.s. fluctuating velocity $U$ (used to compute $Re$) does not accurately characterise the convective velocity at the scale that contributes most to convective heat transport in the rotationally-constrained regime. Alternatively, it may indicate that the dynamics of the CZ are influenced by the heated and cooled regions (by analogy with the influence that boundary layers can have on the bulk in boundary-driven convection; \citealp{2000JFM...407...27G,2012JFM...691..568K,2016JFM...808..690G}). We note that forming a volume-integrated relation between $Re$ and the viscous dissipation rate is one route to obtaining the inertial scaling for $Re$ (Equation \ref{eq:ReMLT}; \citealp{2016JFM...808..690G}). As we are interested in the dynamics integrated over the CZ, this integral relation will not be exact, but instead contain `boundary terms' that describe the exchange of kinetic energy between the CZ and the heated and cooled regions. Understanding the details of this exchange may be important for understanding the scaling behaviour of $Re$ in the CZ.

Several aspects of the flows here are unlike those realised in a star like the Sun. We believe most of these discrepancies result from choices and simplifications made in our modeling (rather than from a direct dependence on diffusivity), but testing this assertion will require future work.  For example, none of the zonal flows in our calculations are particularly `Solar-like'; those that exhibit a fast equator typically also have a `polar vortex', consisting of prograde flow at high latitudes.  This may reflect the narrow aspect ratio of the convective shell considered here (see, e.g., \citealp{1979ApJ...231..284G}). Alternatively, we note previous work has found that the emergence of polar vortices in simulations of shell convection can be dependent on the initial conditions \citep{2014A&A...570A..43K}. More generally, we have restricted ourselves to the simplest case of Boussinesq, hydrodynamic convection. This set-up has the advantage of being easily relatable to theory. However, both magnetism and compressibility have important effects on the dynamics of rotating, spherical shell convection (e.g., \citealp{2012Icar..219..428G,2016GeoJI.204.1120Y,2023A&A...669A..98K,2015ApJ...798...51H,2022ApJ...933..199H,10.1093/mnras/staf1081}). Therefore, incorporating these effects is an obvious and important follow-up to this work. In addition, we have not analysed the sensitivity of our results to the functional form of the prescribed internal heating and cooling profile. We intend to study these effects in future work. 

In spite of these limitations, we argue that our findings have very promising implications for the numerical modeling of convection in stellar (and giant planet) interiors.  All simulations of interior convection operate with diffusivities (whether explicit or numerical) that are orders of magnitude larger than those of real objects, so it is vital to understand how these diffusivities alter the results of the simulations.  Our demonstration here that aspects of the heat transfer and flow dynamics are diffusion-free (i.e., independent of the diffusivities) will, we believe, enable more confident extrapolation from our results to real environments.  To be explicit, we think that if a simulation in the rotationally-uninfluenced or rotationally-influenced regimes could be conducted with vastly lower diffusivities (but at the same value of $Ro_{cv, F}$, and in a otherwise identical setup), its temperature gradient, integrated fluctuating kinetic energy, and differential rotation would not differ from those achieved in our simulations at high supercriticality. Moreover, we believe that some aspects of the rotationally-constrained cases are approaching a diffusion-free state for which a similar conclusion could be drawn, but simulations with a higher supercriticality are required to demonstrate this conclusively.

\begin{acknowledgments}
Data supporting this manuscript has been uplodaded to Zenodo: \href{https://zenodo.org/records/17961728}{zenodo.17961728}. We would like to acknowledge an anonymous referee whose feedback helped us to improve this manuscript throughout. We thank Benjamin Brown and Keaton Burns for useful suggestions with respect to configuring and using Dedalus. NTL and TJ-H benefited from stimulating discussions at the workshop \emph{Rotating Turbulence: Interplay and Separability of Bulk and Boundary Dynamics}, hosted by Institute for Pure and Applied Mathematics (IPAM) at the University of California, Los Angeles. IPAM is supported by the National Science Foundation (grant no. DMS-1925919). MKB thanks the Isaac Newton Institute for Mathematical Sciences, Cambridge, for support and hospitality during the programme \emph{Frontiers in Dynamo Theory: From the Earth to the Stars (DYT2)}, where preliminary work on this topic was conducted; this was supported by EPSRC grant no EP/R014604/1.  MKB also thanks Evan Anders, and other participants in the DYT2 programme, for useful discussions. NTL, TJ-H, MKB, LKC, and ST received funding from the STFC under grant agreements ST/Y002156/1, ST/Y002296/1, and ST/Y002113/1. LKC also gratefully acknowledges support from the STFC under grant agreement ST/X001083/1, and TJ-H was additionally supported by an STFC Studentship under grant agreement ST/W507453/1.  This work used the DiRAC Memory Intensive service (Cosma8) at Durham University, managed by the Institute for Computational Cosmology, and the DiRAC Data Intensive service (DIaL3) at the University of Leicester managed by the University of Leicester Research Computing Service. These facilities are managed on behalf of the STFC DiRAC HPC (www.dirac.ac.uk). The DiRAC services at Durham and Leicester were funded by BEIS, UKRI and STFC capital funding, and STFC operations grants. The service at Durham received funding from Durham University. DiRAC is part of the UKRI Digital Research Infrastructure.  
\end{acknowledgments}

\begin{contribution}
NTL, MKB, LKC, and ST designed the research. NTL performed the numerical simulations and data analysis. All authors contributed to the interpretation of results and to writing the manuscript. 
\end{contribution}

%

\software{Dedalus v3 \citep{2020PhRvR...2b3068B}.}



\appendix

\section{Simulation Input Parameters}\label{ap:A}

\addtolength{\tabcolsep}{0.62em}
\startlongtable
\begin{deluxetable*}{c|rccccc}
\tablewidth{0pt}
\tablecaption{All simulations were run with $Pr=1$ and $\eta=0.8$, where $\eta=r_{\text{i}}/r_{\text{o}}$ is the aspect ratio of the convection zone, and use a non-dimensional depth $\tilde{\delta}=0.2$ for the heated and cooled regions. Below, $N_r$ is the number of Chebyshev modes used, and $\ell_{\text{max}}$ is the maximum spherical harmonic degree. $n_{r}$ and $n_{\theta}$ are the corresponding number of radial and latitudinal grid points (using 3/2 dealiasing). \label{tab:A}}
\tablehead{
\colhead{} & \multicolumn{6}{c}{Input Parameters}  \\[-4pt] 
\colhead{$Ro_{\text{cv,F}}$} & \colhead{$Ra_{\text{F}}$} & \colhead{$Ta$}& \colhead{$N_{\text{r}}$} &\colhead{$\ell_{\text{max}}$} & \colhead{$n_{r}$} & \colhead{$n_{\theta}$}
}
\startdata
$1.44\times10^{-2}$ & $9.5\times10^{7}$ & $1\times10^{9}$  & 139 & 383 & 210 & 576 \\ 
\hline 
$2.12\times10^{-2}$ & $3\times10^{5}$   & $1\times10^{7}$  & 99  & 127 & 150 & 192 \\ 
\hline 
$3.11\times10^{-2}$ & $1.6\times10^{5}$ & $3\times10^{6}$  & 99  & 127 & 150 & 192 \\ 
                    & $9.5\times10^{5}$ & $1\times10^{7}$  & 99  & 127 & 150 & 192 \\ 
                    & $4.9\times10^{6}$ & $3\times10^{7}$  & 119 & 255 & 180 & 384 \\ 
                    & $3\times10^{7}$   & $1\times10^{8}$  & 119 & 255 & 180 & 384 \\
                    & $9.5\times10^{8}$ & $1\times10^{9}$  & 139 & 383 & 210 & 576 \\ 
\hline 
$4.56\times10^{-2}$ & $3\times10^{6}$   & $1\times10^{7}$  & 119 & 255 & 180 & 384 \\ 
\hline 
$6.70\times10^{-2}$ & $3\times10^{5}$   & $1\times10^{6}$  & 99  & 127 & 150 & 192 \\ 
                    & $1.6\times10^{6}$ & $3\times10^{6}$  & 99  & 127 & 150 & 192 \\ 
                    & $9.5\times10^{6}$ & $1\times10^{7}$  & 119 & 255 & 180 & 384 \\ 
                    & $4.9\times10^{7}$ & $3\times10^{7}$  & 119 & 255 & 180 & 384 \\ 
                    & $3\times10^{8}$   & $1\times10^{8}$  & 139 & 383 & 210 & 576 \\ 
\hline  
$9.83\times10^{-2}$ & $9.5\times10^{5}$ & $1\times10^{6}$  & 99  & 127 & 150 & 192 \\ 
                    & $3\times10^{7}$   & $1\times10^{7}$  & 119 & 255 & 180 & 384 \\ 
\hline 
$1.44\times10^{-1}$ & $3\times10^{6}$   & $1\times10^{6}$  & 99  & 127 & 150 & 192 \\ 
                    & $9.5\times10^{7}$ & $1\times10^{7}$  & 139 & 383 & 210 & 576 \\ 
\hline 
$2.12\times10^{-1}$ & $9.5\times10^{6}$ & $1\times10^{6}$  & 99  & 127 & 150 & 192 \\ 
                    & $3\times10^{8}$   & $1\times10^{7}$  & 139 & 383 & 210 & 576 \\ 
\hline 
$3.11\times10^{-1}$ & $9.5\times10^{5}$ & $1\times10^{5}$  & 79  & 63  & 120 & 96  \\
                    & $4.9\times10^{6}$ & $3\times10^{5}$  & 99  & 127 & 150 & 192 \\
                    & $3\times10^{7}$   & $1\times10^{6}$  & 119 & 255 & 180 & 384 \\ 
                    & $1.6\times10^{8}$ & $3\times10^{6}$  & 139 & 383 & 210 & 576 \\ 
                    & $9.5\times10^{8}$ & $1\times10^{7}$  & 159 & 511 & 240 & 768 \\ 
\hline 
$4.56\times10^{-1}$ & $3\times10^{6}$   & $1\times10^{5}$  & 99  & 127 & 180 & 384 \\ 
                    & $9.5\times10^{7}$ & $1\times10^{6}$  & 139 & 383 & 210 & 576 \\ 
                    & $3\times10^{9}$   & $1\times10^{7}$  & 159 & 511 & 240 & 768 \\ 
\hline  
$6.70\times10^{-1}$ & $9.5\times10^{6}$ & $1\times10^{5}$  & 119 & 255 & 180 & 384 \\ 
                    & $4.9\times10^{7}$ & $3\times10^{5}$  & 139 & 383 & 210 & 576 \\ 
                    & $3\times10^{8}$   & $1\times10^{6}$  & 139 & 383 & 210 & 576 \\ 
\hline 
$9.83\times10^{-1}$ & $3\times10^{7}$   & $1\times10^{5}$  & 119 & 255 & 180 & 384 \\ 
                    & $9.5\times10^{8}$ & $1\times10^{6}$  & 159 & 511 & 240 & 768 \\ 
\hline 
$1.44$              & $9.5\times10^{7}$ & $1\times10^{5}$  & 139 & 383 & 210 & 576 \\ 
                    & $3\times10^{9}$   & $1\times10^{6}$  & 159 & 511 & 240 & 768  \\ 
\hline 
$2.12$              & $3\times10^{8}$   & $1\times10^{5}$  & 159 & 511 & 240 & 768  \\ 
                    & $9.5\times10^{9}$ & $1\times10^{6}$  & 159 & 511 & 240 & 768  \\ 
\hline  
$3.11$              & $9.5\times10^{8}$ & $1\times10^{5}$  & 159 & 511 & 240 & 768  \\ 
\hline  
$4.56$              & $3\times10^{9}$   & $1\times10^{5}$  & 159 & 511 & 240 & 768  \\
\enddata
\end{deluxetable*}

\bibliography{main}{}
\bibliographystyle{aasjournalv7}

\end{document}